\documentclass[aps,twocolumn,prl,superscriptaddress,amsmath,amssymb]{revtex4-1}
\newcommand{\kets}[1]{\lvert#1\rangle}

\newcommand{\mean}[1]{\left<#1\right>}
\newcommand{\means}[1]{\langle#1\rangle}

\ifx\pdfoutput\undefined
\usepackage[dvipdfmx]{graphicx}
\usepackage[dvipdfmx]{hyperref}
\else
\usepackage{graphicx}
\usepackage{hyperref}
\fi

\usepackage{txfonts}
\usepackage[T1]{fontenc}
\usepackage{xspace}

\usepackage{dcolumn}
\usepackage{bm}
\usepackage{amsmath}
\usepackage{booktabs}
\usepackage[usenames]{color}
\usepackage{xcolor}

\begin{document}
\let\emph\textit

\title{
 Successive Majorana Topological Transitions Driven by a Magnetic Field in the Kitaev Model
}

\author{Joji Nasu}
\affiliation{
  Department of Physics, Tokyo Institute of Technology,
  Meguro, Tokyo 152- 8551, Japan
}
\author{Yasuyuki Kato}
\affiliation{
  Department of Applied Physics, University of Tokyo,
  Bunkyo, Tokyo 113-8656, Japan
}
\author{Yoshitomo Kamiya}
\affiliation{
  Condensed Matter Theory Laboratory, RIKEN,
  Wako, Saitama 351-0198, Japan
}
\author{Yukitoshi Motome}
\affiliation{
  Department of Applied Physics, University of Tokyo,
  Bunkyo, Tokyo 113-8656, Japan
}

\date{\today}
\begin{abstract}
 We study quantum phase transitions in the honeycomb Kitaev model under a magnetic field, focusing on the topological nature of Majorana fermion excitations.
 We find a gapless phase between the low-field gapless quantum spin liquid and the high-field gapped forced-ferromagnetic state for the antiferromagnetic Kitaev model in the [001] field by using the Majorana mean-field theory, in conjunction with the exact diagonalization and the spin-wave theory supporting the validity of this approach.
The transition between the two gapless phases is driven by a topological change of the Majorana spectrum --- line node formation interconnecting two Majorana cones.
The peculiar change of the Majorana band topology is rationalized by a sign 
change of the effective Kitaev coupling by the magnetic field, which does not occur in the ferromagnetic Kitaev case.
 Upon tilting the magnetic field away from [001], the two gapless phases become gapped and topologically nontrivial, characterized by nonzero Chern numbers with different signs.
 The sign change of the Chern number leads to a reversal of the thermal edge current in the half-quantized thermal Hall effect.
\end{abstract}

\maketitle

 The concept of topology plays a central role in the current forefront of condensed matter physics.
 This holds particularly true in the study of quantum spin liquids (QSLs), where any kind of conventional order is suppressed by quantum fluctuations~\cite{ISI:000275366100033,Savary2017}.
 Originating from their topological order, QSLs can host fractionalized excitations~\cite{PhysRevLett.96.060601}, which may be observed as excitation continua in dynamical spin responses and low-temperature asymptotic behaviors of specific heat and thermal conductivity~\cite{han2012fractionalized,yamashita2008thermodynamic,ISI:000278318600025,PhysRevLett.112.177201,watanabe2016emergence}.
 In addition, fractionalized excitations can form a nontrivial topological band structure.
 To observe this, thermal Hall measurements have been performed in QSL candidate materials in an applied magnetic field~\cite{watanabe2016emergence,Kasahara2017pre,Hentrich2018pre}.

 Such topological nature and fractionalized excitations have been actively debated in the past decade for Kitaev-type QSLs.
 The Kitaev model has an exact QSL ground state associated with Majorana fermions emergent from spin fractionalization~\cite{Kitaev2006,PhysRevLett.98.247201,PhysRevLett.102.017205}.
 Although most of the candidate materials exhibit a magnetic order at low temperature~\cite{PhysRevB.82.064412,PhysRevLett.108.127203,PhysRevB.90.041112,PhysRevB.91.094422,PhysRevB.91.144420,Majumder2015,Kitagawa2018nature}, their excitation spectra observed by neutron and Raman scatterings~\cite{PhysRevLett.114.147201,banerjee2016proximate,Nasu2016nphys,Banerjee2017} and longitudinal thermal conductivity~\cite{hirobe2017} above N\'eel temperature show anomalous features likely related to Majorana fermions.

 Recently, the magnetic-field effect suppressing the magnetic order became a topical issue in the Kitaev candidate materials~\cite{Johnson2015,Wolter2017,Baek2017,Sears2017,Ponomaryov2017,Wang2017,Zheng2017,Leahy2017,banerjee2018excitations,Yu2018,Hentrich2018,janvsa2017pre}.
 Theoretically, a weak magnetic field in the perturbative regime is known to open a gap in the Majorana fermion spectrum, which induces the half-quantized thermal Hall conductivity due to the Majorana chiral edge mode~\cite{Kitaev2006,Nasu2017}.
 Interestingly, a thermal Hall effect was observed in the Kitaev candidate material $\alpha$-RuCl$_3$ in a magnetic field above the N\'eel temperature~\cite{Kasahara2017pre,Hentrich2018pre}.
 However, a magnetic field beyond the perturbative treatment violates the exact solvability of the Kitaev model.
 So far, the magnetic field effect has been studied theoretically by numerical calculations for finite-size clusters~\cite{Jiang2011,yadav2016kitaev,Zhu2017pre,Gohlke2018pre}, the slave-particle mean-field (MF) theory combined with variational Monte Carlo calculations~\cite{Liu2017pre}, and the linear spin-wave (SW) approximation~\cite{Janssen2017,mcclarty2018pre,Joshi2018pre}.
 While these methods are versatile and widely used for frustrated spin systems, they are not so straightforward to address the aspect associated with fractional Majorana quasiparticles, which is the most significant characteristics of the Kitaev QSL.
 Moreover, previous studies mostly focused on the magnetic field applied in the [111] direction.
 It remains a theoretical challenge to provide a comprehensive understanding of field-induced phenomena beyond the perturbative treatment by Kitaev, especially from the viewpoint of Majorana fermions.

 In this Letter, we investigate the effect of the magnetic field on the Kitaev QSL, focusing on the field along the [001] and its proximate directions.
 The magnetic-field dependence is examined by using the exact diagonalization (ED), the Majorana MF theory, and the SW theory.
 We find that, in the Majorana MF solution, which well reproduces the magnetization and spin correlations obtained by the ED, the [001] field induces two successive topological transitions in the antiferromagnetic (AFM) Kitaev model:
 One is from the low-field gapless Kitaev QSL to a newly-found gapless intermediate phase, and the other is from the intermediate phase to the high-field forced-ferromagnetic (FF) state.
 This is in stark contrast to the ferromagnetic (FM) case, which shows a direct transition from the QSL to the FF state.
 Analyzing the low-energy Majorana spectrum, we clarify that the intermediate phase appears as a consequence of a topological change of the Majorana spectrum in momentum space.
This is understood by a continuous change of the effective Kitaev coupling for the spin $z$ component from AFM to FM by the [001] field.
 Upon tilting the field, the low-field QSL and the intermediate state are gapped out.
 We show that the resulting phases are characterized by nonzero Chern numbers with opposite signs, and hence, they can be distinguished via the sign reversal of the half-quantized thermal Hall coefficients.

 The Hamiltonian for the Kitaev model under the magnetic field is given by
\begin{eqnarray}
  {\cal H}=-\sum_\gamma\sum_{\means{jj'}_\gamma}J_\gamma S_j^\gamma S_{j'}^\gamma - \bm{h} \cdot \sum_{j} \bm{S}_j,
  \label{eq:1}
\end{eqnarray}
where $S_j^\gamma$ is the $\gamma$ component of an $S=1/2$ spin at site $j$ on a honeycomb lattice, $\means{jj'}_\gamma$ denotes nearest neighbors on the $\gamma$ bond ($\gamma=x,y,z$), which corresponds to one of the three bond orientations, and $\bm{h} = (h_x,h_y,h_z)$ is the magnetic field.
 We consider the isotropic case with $J_x=J_y=J_z=J$.

\begin{figure}[t]
 \begin{center}
  \includegraphics[width=\columnwidth,clip]{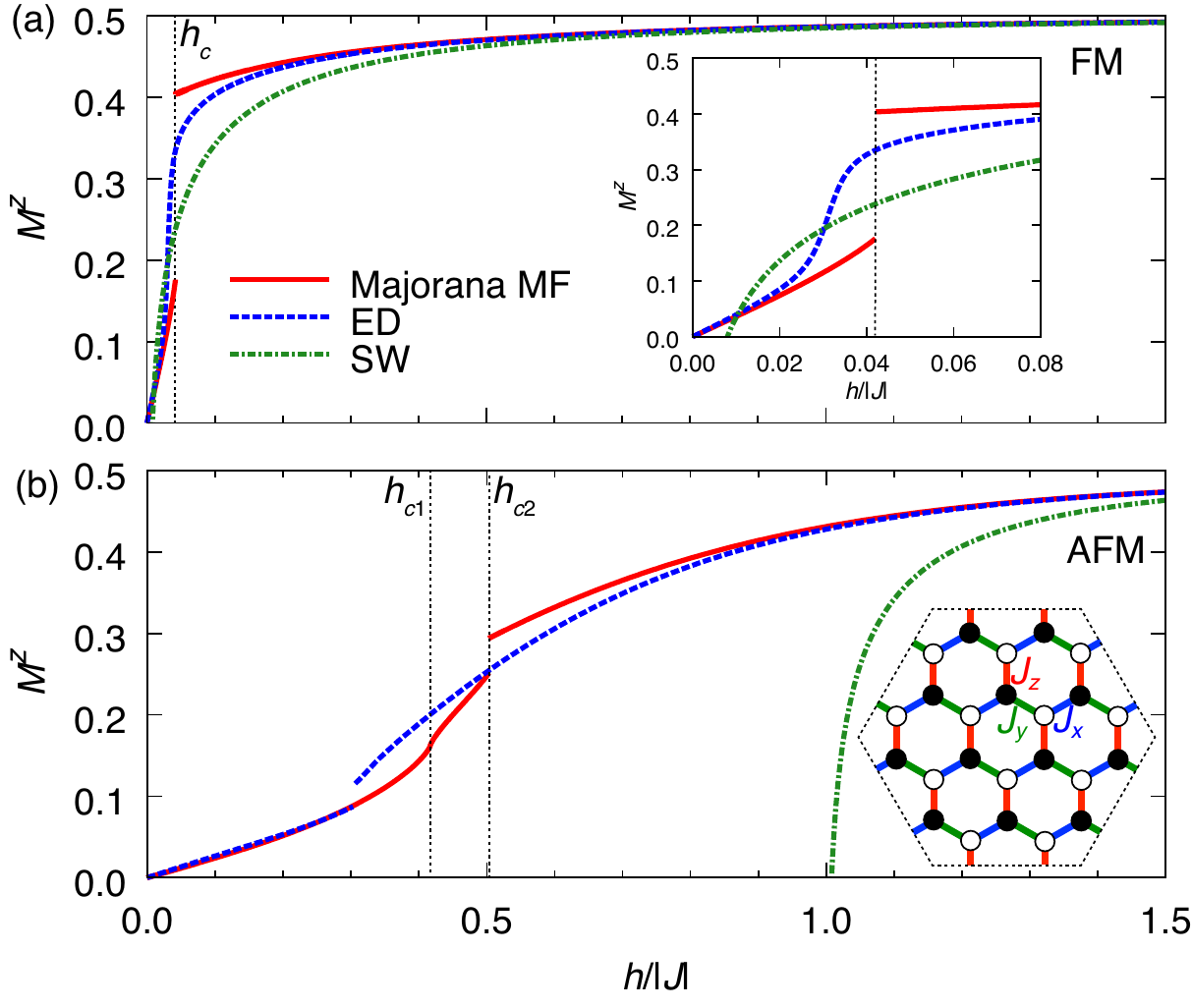}
  \caption{
 Magnetization curve for (a) the FM Kitaev model and (b) the AFM one in the [001] field, which are calculated by the ED in the 24-site cluster shown in the inset of (b), the Majorana MF theory, and the SW theory.
 The enlarged view of the low-field region in the FM case is shown in the inset of (a).
 $h_c$, $h_{c1}$, and $h_{c2}$ are the critical fields where the Majorana MF results show anomalies.
}
  \label{mag}
 \end{center}
\end{figure}

 Let us begin with the magnetic field in  the [001] direction, i.e., $h_x=h_y=0$ and $h_z=h$, by using the ED, the Majorana MF theory, and the SW theory.
 The ED calculation is performed for the 24-site cluster shown in the inset of  Fig.~\ref{mag}(b), which has been widely used to study the ground state of extended Kitaev models~\cite{PhysRevLett.105.027204,PhysRevLett.110.097204,PhysRevLett.112.077204,Okamoto2013,PhysRevLett.113.107201,PhysRevB.93.174425}.
 On the other hand, our Majorana MF theory is based on the Jordan-Wigner transformation~\cite{PhysRevB.76.193101,PhysRevLett.98.087204,1751-8121-41-7-075001,PhysRevLett.113.197205}, which fermionizes the Hamiltonian in Eq.~\eqref{eq:1}.
 This formalism has an advantage that it does not require local constraints, which are usually difficult to impose exactly at the MF level.
 We obtain ${\cal H}=-iJ\sum_{\gamma=x,y}\sum_{\mean{jj'}_\gamma}a_j b_{j'}-J\sum_{\mean{jj'}_z}i a_j b_{j'}i \bar{a}_j \bar{b}_{j'}-ih\sum_j\left(a_j\bar{a}_j-b_j\bar{b}_j\right)$,
 where $a_j$ and $\bar{a}_j$ ($b_j$ and $\bar{b}_j$) are the Majorana fermion operators for black (white) sites shown in the inset of Fig.~\ref{mag}(b)~\cite{suppl}.
In this representation, the $z$ bond terms act as interactions between Majorana fermions.
 At $h=0$, this model can be solved exactly as $i \bar{a}_j \bar{b}_{j'}$ on each $z$ bond $\mean{jj'}_z$ is a $Z_2$ conserved quantity.
 For $h \ne 0$, however, this no longer holds; thus, we apply the MF decoupling to the interactions on the $z$ bonds and obtain the MF Hamiltonian ${\cal H}^{\rm MF}=\sum_{\bm{k}}' \bm{c}_{\bm{k}}^\dagger {\cal H}_{\bm{k}}^{\rm MF} \bm{c}_{\bm{k}}$ with $\bm{c}_{\bm{k}}=\left(a_{\bm{k}},b_{\bm{k}},\bar{a}_{\bm{k}},\bar{b}_{\bm{k}}\right)^{\rm T}$, where ${\cal H}_{\bm{k}}^{\rm MF}$ is a $4\times 4$ matrix and the sum of the momentum $\bm{k}$ runs over the half of the Brillouin zone so as to avoid the redundancy between $\bm{k}$ and $-\bm{k}$.
 Finally, the present SW theory is a standard linear SW approximation, which is expected to correctly describe the behavior near the fully polarized state in high fields.
 Further details of the latter two methods are given in Supplemental Material (SM)~\cite{suppl}.

 Figures~\ref{mag}(a) and~\ref{mag}(b) show the magnetization $M^z=\means{S_{j}^z}$ in the [001] field for $J>0$ (FM) and $J<0$ (AFM), respectively.
 In both cases, the results obtained by the Majorana MF theory show a similar $h$ dependence to those by the ED.
 The agreement is also seen for the spin correlations~\cite{suppl}.
  This is not the case for the SW theory at low fields; the result deviates considerably from the ED and MF ones, especially for the AFM case, whereas the deviations are small near the saturation as expected.
  The comparison demonstrates the efficiency of our Majorana MF approach not only in the high-field FF phase but also in low-field region including the QSL phase.

\begin{figure}[b]
 \begin{center}
  \includegraphics[width=\columnwidth,clip]{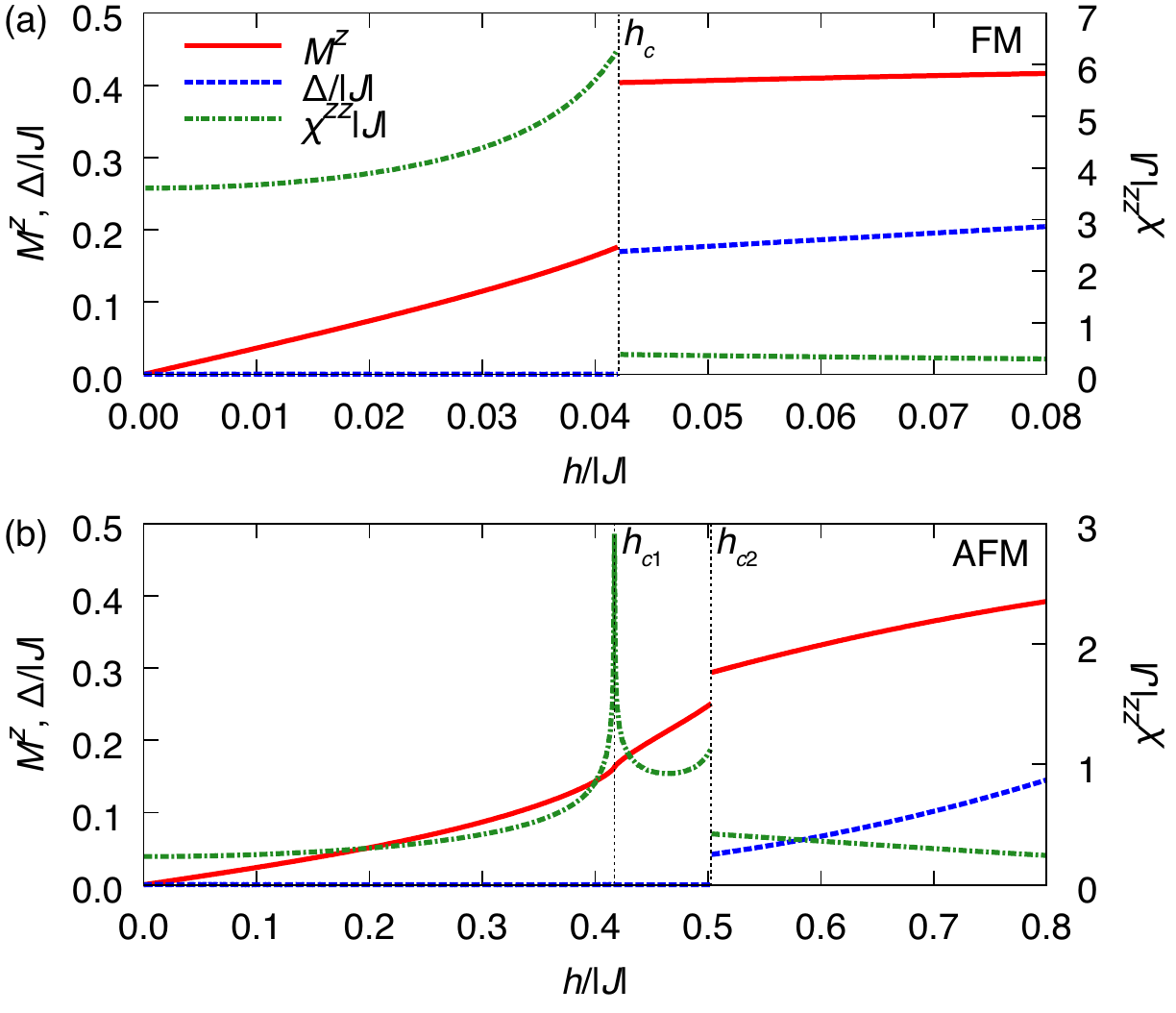}
  \caption{
 Majorana MF results for the [001] field dependence of the magnetization $M^z$, the Majorana gap $\Delta$, and the magnetic susceptibility $\chi^{zz}$ for (a) the FM Kitaev model and (b) the AFM one.
}
  \label{phys}
 \end{center}
\end{figure}

 In the FM case, $M^z$ obtained by the MF theory exhibits a jump at $h_c/|J|\simeq 0.042$, indicating a first-order phase transition, while the ED result shows an abrupt change around $h_c$; see the inset of Fig.~\ref{mag}(a).
 Correspondingly, the magnetic susceptibility $\chi^{zz}\equiv dM^z/dh$ obtained by the MF theory changes discontinuously at $h_c$, as shown in Fig.~\ref{phys}(a).
 For $h>h_c$, a nonzero gap $\Delta$ opens in the Majorana fermion spectrum.
 Meanwhile, in the AFM case, the MF result shows two singularities in $M^z$, a continuous one at $h_{c1}/|J|\simeq 0.417$ and a discontinuous one at $h_{c2}/|J|\simeq 0.503$, while the ED shows a single discontinuity at $h/ \lvert{J}\rvert \simeq 0.305$ [Fig.~\ref{mag}(b)].
 As shown in Fig.~\ref{phys}(b), $\chi^{zz}$ obtained by the MF theory diverges in the continuous transition at $h_{c1}$, while it jumps at $h_{c2}$.
 The Majorana gap remains zero up to $h_{c2}$, indicating that the intermediate phase for $h_{c1}<h<h_{c2}$ is also gapless.
 Leaving the apparent difference between the MF and ED results for later discussion, below we first analyze the nature and the origin of the intermediate phase obtained by the Majorana MF theory in the AFM case.

\begin{figure*}[t]
 \begin{center}
  \includegraphics[width=2\columnwidth,clip]{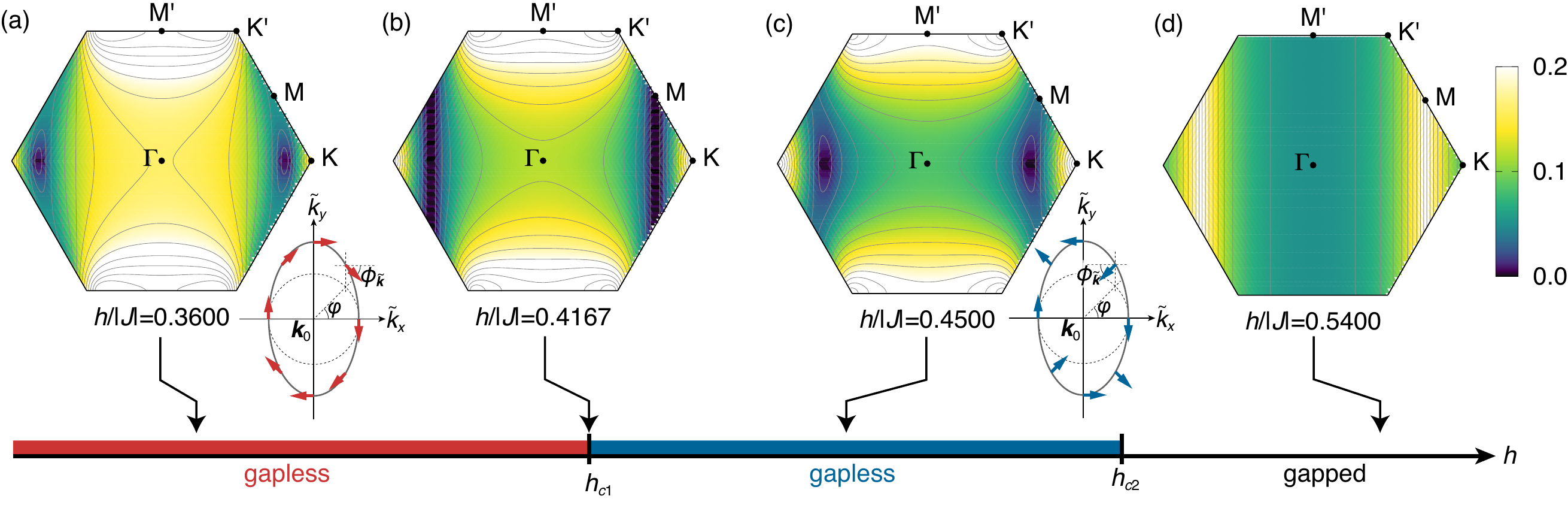}
  \caption{
Lower-energy Majorana dispersions in the Majorana MF solutions for the AFM Kitaev model under the [001] magnetic field $h$, with the corresponding phases indicated at the bottom along the $h$-axis.
The dispersion relations along the symmetric line are shown in Fig.~S1 of SM~\cite{suppl}.
In the lower-right panels of (a) and (c), the phase $\phi_{\tilde{\bm{k}}}$ of the low-energy eigenfunction in Eq.~(\ref{eq:wf}) is schematically shown by arrows along the elliptical constant-energy line parameterized by $\varphi$ around the nodal point at $\bm{k}_0=(k_0,0)$ with $k_0>0$.
}
  \label{disp}
 \end{center}
\end{figure*}

\begin{figure}[t]
 \begin{center}
  \includegraphics[width=\columnwidth,clip]{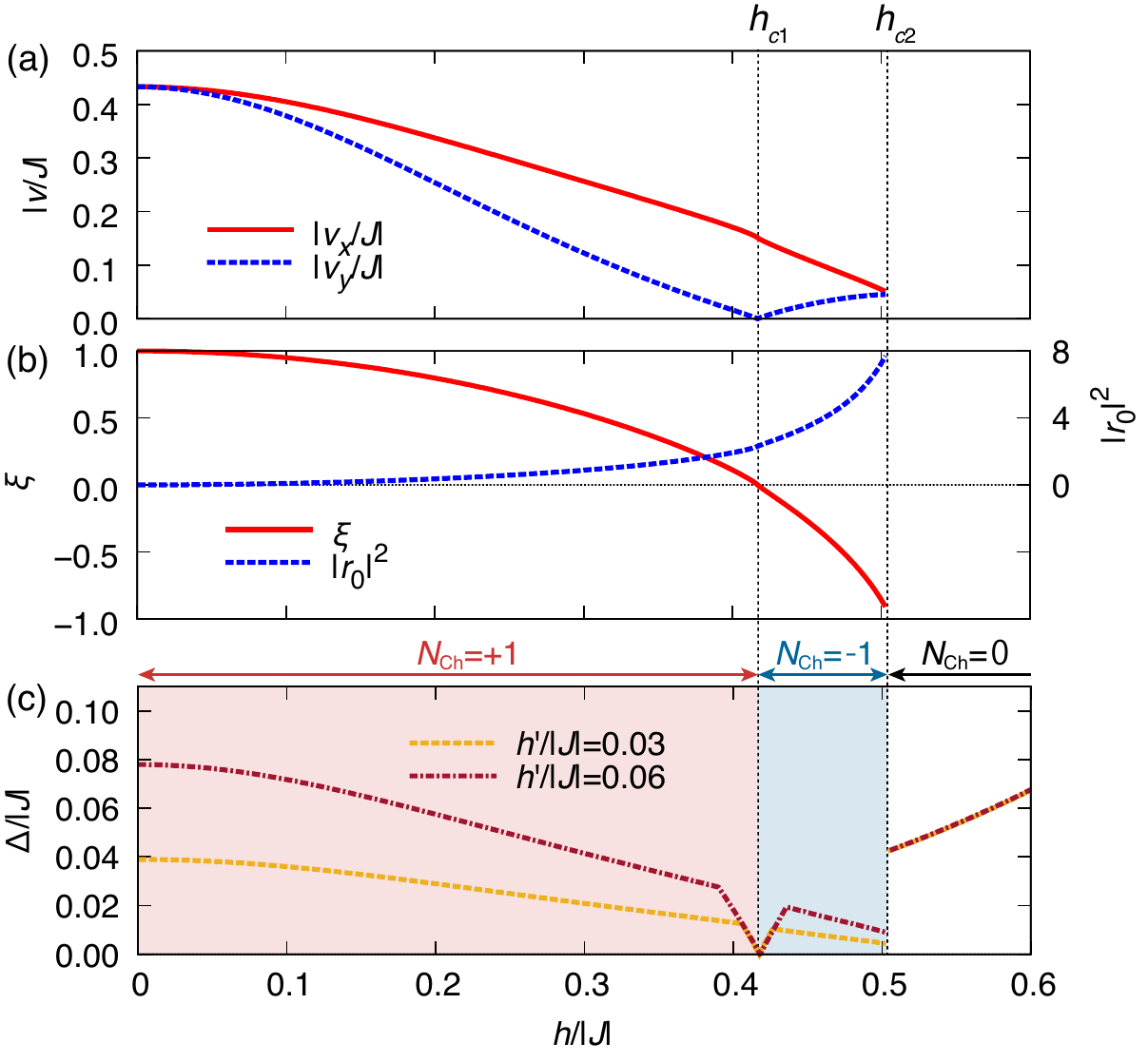}
  \caption{
  Magnetic field dependence of (a) the velocities $|v_x|$ and $|v_y|$ for the low-energy dispersion and (b) the parameters $\xi$ and $|r_0|^2$, which characterize the low-energy dispersion and the eigenfunction, respectively, in the AFM Kitaev model.
(c) Majorana gap under the effective magnetic field $h'$.
 The Chern numbers $N_{\rm Ch}$ calculated for $h'>0$ by the MF theory are given on the top of (c).
}
  \label{lowene}
 \end{center}
\end{figure}

 The evolution of the low-energy Majorana dispersions for the AFM case while increasing $h$ is shown in Fig.~\ref{disp}.
 At $h=0$, there are two Majorana cones at the K and K' points in the Brillouin zone with $C_3$ rotational symmetry around the $\Gamma$ point.
 A nonzero $h$ breaks this symmetry, while the mirror symmetries with respect to the vertical and horizontal lines crossing the $\Gamma$ point are preserved.
 While increasing $h$, the Majorana cone at the K point moves horizontally towards the $\Gamma$ point; see Fig.~\ref{disp}(a) (the cone at the K' point moves in parallel towards the $\Gamma$ point in the adjacent Brillouin zone).
 While these nodal points move away from the K and K' points, the velocity in the $y$ direction becomes smaller, which eventually vanishes at $h=h_{c1}$.
 Consequently, there appears a vertical line node passing through the M point, connecting the Majorana cones [Fig.~\ref{disp}(b)].
 For $h_{c1} < h < h_{c2}$, a pair of Majorana cones reappear on the $\Gamma$--K and $\Gamma$--K' horizontal lines, respectively, drifting further towards the $\Gamma$ points [Fig.~\ref{disp}(c)].
 This drift is terminated at $h_{c2}$, above which the Majorana dispersion is gapped [Fig.~\ref{disp}(d)].
 In the case of the FM Kitaev model, where the intermediate phase is absent, a pair of Majorana cones drift similarly as $h \to h_c$, but are simply gapped out for $h > h_c$.

 Analyzing the Majorana eigenstates, we find that the continuous transition with the line node formation at $h = h_{c1}$ in the AFM case is associated with the change of the topological nature of the Majorana fermions.
 When the Majorana nodal point is located at $\bm{k}_0=(k_0,0)$ on the $\Gamma$--K line, the dispersion relation in the low-energy limit is given by
$
E_{\tilde{\bm{k}}}\sim\sqrt{|v_x|^2 \tilde{k}_x^2+|v_y|^2 \tilde{k}_y^2}
$,
where $v_x$ and $v_y$ are the velocities in the $x$ and $y$ directions, respectively, and $\tilde{\bm{k}}=\bm{k}-\bm{k}_0$.
As shown in Fig.~\ref{lowene}(a), the velocities $v_x$ and $v_y$ becomes anisotropic, and $v_y$ vanishes at $h_{c1}$, corresponding to the vertical line node formation~\cite{suppl}.

 The eigenfunction for the energy $E_{\tilde{\bm{k}}}$ around $\bm{k}_0$ is obtained as
\begin{eqnarray}
\kets{\varphi_{\tilde{\bm{k}}}} \sim
\frac{1}{\sqrt{1+|r_{\tilde{\bm{k}}}|^2}}
\left(1,e^{i\phi_{\tilde{\bm{k}}}-i\delta_{\bm{k}}},
r_{\tilde{\bm{k}}}e^{i\phi_{\tilde{\bm{k}}}},
r_{\tilde{\bm{k}}}e^{-i\delta_{\bm{k}}}
\right)^\mathrm{T},
\label{eq:wf}
\end{eqnarray}
for the basis of ($a_{\bm{k}},b_{\bm{k}},\bar{a}_{\bm{k}},\bar{b}_{\bm{k}}$) with $e^{-i\phi_{\tilde{\bm{k}}}}=(v_y \tilde{k}_y +i v_x \tilde{k}_x)/E_{\tilde{\bm{k}}}$  and $\delta_{\bm{k}}=k_y/\sqrt{3}$, where we take the lengths of the primitive translation vectors unity~\cite{suppl}.
 The norm square of the complex parameter $r_0 \equiv r_{\tilde{\bm{k}}=0}$ varies with $h$ as shown in Fig.~\ref{lowene}(b).
 At $h=0$, $|r_{0}|^2$ vanishes because the low-energy excitation is governed only by $(a_j,b_{j})$.
 While increasing $h$, $|r_{0}|^2$ monotonically increases corresponding to the mixing between $(a_j,b_{j})$ and $(\bar{a}_j,\bar{b}_{j})$.
 In Eq.~\eqref{eq:wf}, the phase $\phi_{\tilde{\bm{k}}}$ plays an essential role for determining the topological nature of the low-energy Majorana fermions.
 The contour of the low-energy dispersion around $\tilde{\bm{k}}_0$ is an ellipse parameterized by $\varphi$ as
$(\tilde{k}_x,\tilde{k}_y)=\left(\frac{E_{\tilde{\bm{k}}}}{|v_x|}\cos\varphi,
\frac{E_{\tilde{\bm{k}}}}{|v_y|}\sin\varphi\right)$.
The phase $\varphi$ of the ellipse is related with $\phi_{\tilde{\bm{k}}}$ as $\phi_{\tilde{\bm{k}}}={\rm sgn}(k_0\xi)\varphi+\left(n+\frac{1}{2}\right)\pi$ with an integer $n$ [see the lower-right panels of Figs.~\ref{disp}(a) and~\ref{disp}(c)], where $\xi = -2 \cos (k_0/2)$.
 This gives a correspondence between the quantum phase of the wavefunction and the momentum on the energy contour surrounding each Majorana nodal point.
 As shown in Fig.~\ref{lowene}(b), $\xi$ monotonically decreases from $1$ as $h$ increases and changes the sign at $h = h_{c1}$, which results in the reversal of the chirality of $\phi_{\tilde{\bm{k}}}$ around the nodal point.
 This indicates that the topological nature of the Majorana fermions changes through this transition with the line node formation.

 The topological transition manifests itself more evidently when the Majorana cones are gapped out by tilting the field from [001].
 Here, we show this by perturbing the Hamiltonian with a symmetry-allowed term in the presence of small tilting,
 ${\cal H}'_h=-h'\sum_{[jj'j'']_{\gamma\gamma''}}S_j^{\gamma}S_{j'}^{\gamma'}S_{j''}^{\gamma''}$~\cite{Kitaev2006}, where $[jj'j'']_{\gamma\gamma''}$ represents neighboring three sites connected by consecutive $\gamma$ bond $\langle{jj'}\rangle_\gamma$ and $\gamma''$ bond $\langle{j'j''}\rangle_{\gamma''}$ and $\gamma'$ denotes the index of the other bond connected to the site $j'$ (i.e., $\gamma' \ne \gamma,\gamma''$).
 For the gapless states below $h_{c2}$, we find that a nonzero $h'$ opens a gap in the Majorana spectrum, and the gap monotonically increases as increasing $h'$, except for $h_{c1}$, as shown in Fig.~\ref{lowene}(c) (note that $h_{c1}$ is slightly shifted by $h'$).
 At this point, the nodal line for $h' = 0$ shrinks to a nodal point at the M point (see Fig.~S2 of SM~\cite{suppl}).
 Meanwhile, in the gapped phases on both sides of $h_{c1}$, we find that the Chern number $N_{\rm Ch}$ is nonzero; the values obtained by the MF theory are shown in Fig.~\ref{lowene}(c).
 At $h=0$, a small positive $h'$ results in $N_{\rm Ch}=+1$ as shown by Kitaev~\cite{Kitaev2006}, which persists up to $h_{c1}$, and it changes into $-1$ for $h_{c1}<h<h_{c2}$.

 The topological change at $h_{c1}$ can be explained by a modulation of the effective Kitaev coupling by the magnetic field.
 The magnetic field along the $z$ direction suppresses the AFM correlation on the $z$ bonds, which can be regarded as the reduction of the effective value of $-J_z$.
 Indeed, the field-evolution of the nodal points of the Majorana cones and that of the low-energy wavefunction for $(a_j,b_{j})$ can be reproduced by the Kitaev model at zero field with a continuous change of $J_z$ from positive to  negative while keeping $J_x=J_y$~\cite{suppl}.
 This suggests that the topological transition induced by $h$
 can be explained by a continuous change of the \emph{effective} interaction $J_z^{\rm eff}$ on the $z$ bond from AFM to FM.
 The reduction of the AFM \emph{effective} interaction is implied also from the weak $h$ approach; the FM interaction $\propto -(h^2/J) \sum_{\means{jj'}_z} S_j^z S_{j'}^z$ is obtained by the second-order perturbation for $h$.
 These observations explain why the topological transition at $h_{c1}$ occurs only in the AFM Kitaev model but not in the FM case, because $J_z^{\rm eff}$ in the latter case is always FM.

 The sign change of $J_z^{\rm eff}$ is shared by the 24-site ED calculation; the $h$ dependence of the spin correlation $\means{S_j^z S_{j'}^z}$ on the $z$ bond calculated by the ED agrees well with the MF result, both showing the sign change induced by $h$~\cite{suppl}.
 Thus, while the present ED lacks a direct indication of the topological transition presumably due to finite-size effects, it supports the scenario of the Majorana MF theory.

 The topological transition presented here could be captured by measuring the thermal Hall effect.
 In a weakly tilted field away from [001], the thermal Hall coefficient $\kappa^{xy}$ is expected to be quantized as $\kappa^{xy}/T=\pi N_{\rm Ch}/12$ in the zero temperature ($T$) limit~\cite{Nomura2012,doi:10.7566/JPSJ.82.023602}, which is a half of that in the conventional Chern insulators.
 Hence, our results indicate that the half-quantized value of $\kappa^{xy}/T$ changes its sign at the topological transition at $h_{c1}$.
 Although the candidate Kitaev materials known thus far appear to have a dominant FM Kitaev coupling, our finding would stimulate the exploration of AFM counterpart.

 In summary, we studied the Kitaev model in the [001] magnetic field.
 We found a gapless phase between the low-field QSL and the high-field FF state in the AFM Kitaev model in the Majorana MF solution.
 We also provided the results of ED and SW theory supporting the validity of the MF theory.
 The new phase appears by a topological transition via the line node formation in the Majorana spectrum where the two Majorana cones are interconnected.
 This topological transition is understood by a continuous change of the effective Kitaev coupling on the $z$ bond from AFM to FM caused by the competition between the AFM Kitaev interaction and the increasing magnetic field.
 When the Majorana nodal points are gapped out by tilting the magnetic field, the Chern number changes its sign between the two phases, which can be observed as a sign reversal of the half-quantized thermal Hall coefficient.
 The present results suggest the possibility of directional switching of the chiral edge mode without inverting the magnetic field, which opens a paradigm of the topological changes in quantum spin systems.
 We note that an intermediate phase was also found for the [111] field in the AFM Kitaev model~\cite{Zhu2017pre,Gohlke2018pre}.
It remains a future issue to clarify the relation to our finding by elucidating the whole magnetic phase diagram.

\begin{acknowledgments}
The authors thank K. Shiozaki for helpful discussions.
This work is supported by Grant-in-Aid for Scientific Research under Grant No. JP15K13533, JP16K17747, JP16H02206, JP18H04223, and JP18K03447.
Parts of the numerical calculations were performed in the supercomputing systems in ISSP, the University of Tokyo.
\end{acknowledgments}

\bibliography{refs}

\clearpage
\onecolumngrid

\appendix
\vspace{15pt}
\begin{center}
{\large \bf ---Supplemental Material---}
\end{center}

\setcounter{figure}{0}
\setcounter{equation}{0}
\setcounter{table}{0}
\renewcommand{\thefigure}{S\arabic{figure}}
\renewcommand{\theequation}{S\arabic{equation}}
\renewcommand{\thetable}{S\Roman{table}}
\baselineskip=6mm

\section{Details of Majorana mean-field theory}

 In this section, we present the details of the Majorana mean-field (MF) theory for the Kitaev model in the [001] magnetic field.
 Using the Jordan-Wigner transformation~\cite{PhysRevB.76.193101,PhysRevLett.98.087204,1751-8121-41-7-075001,PhysRevLett.113.197205}, the spin operators are represented by the Majorana fermions ($a_j,b_j,\bar{a}_i,\bar{b}_j$) as
\begin{align}
 S_j^x=\frac{a_j}{2}\prod_{j'<j}\left(-2S_{j'}^z\right),\quad S_j^y=-\frac{\bar{a}_j}{2}\prod_{j'<j}\left(-2S_{j'}^z\right),\quad S_j^z=\frac{i}{2}a_j\bar{a}_j,
\end{align}
 for the A-sublattice sites and
\begin{align}
S_j^x=\frac{\bar{b}_j}{2}\prod_{j'<j}\left(-2S_{j'}^z\right),\quad S_j^y=-\frac{b_j}{2}\prod_{j'<j}\left(-2S_{j'}^z\right),\quad S_j^z=\frac{i}{2}\bar{b}_j b_j,
\end{align}
for the B-sublattice sites [the A and B sublattice sites are shown by the black and white circles in the inset of Fig.~1(b) of the main text].
 Then, the original spin Hamiltonian in Eq.~(1) of the main text with $J_x=J_y=J_z=J$ and $\bm{h} = (0,0,h)$ is rewritten as
\begin{align}
 {\cal H}=-iJ\sum_{\gamma=x,y}\sum_{\mean{jj'}_\gamma}a_j b_{j'}-J\sum_{\mean{jj'}_z}i a_j b_{j'}i \bar{a}_j \bar{b}_{j'}-ih\sum_j\left(a_j\bar{a}_j-b_j\bar{b}_j\right).
 \label{eq:H_Maj_S}
\end{align}
(The expression is shown in the main text.)
 In the case of $h=0$, $\eta_{\rho}=i\bar{a}_j \bar{b}_{j'}$ is a local conserved quantity taking $\pm1$ defined on each $z$ bond $\rho(=\means{jj'}_z)$, and the model is regarded as a free Majorana fermion system composed of $a_j$ and $b_{j}$.
 However, the magnetic field $h$ gives rise to the hybridization between ($a_j$, $b_j$) and ($b_j$, $\bar{b}_j$) through the last term in Eq.~(\ref{eq:H_Maj_S}), and therefore, $\eta_{\rho}$ is no longer the local conserved quantity for a nonzero $h$.
 In this Majorana representation, the Kitaev interactions on the $z$ bonds are regarded as interactions between Majorana fermions whereas the other terms are given by bilinear forms.
We apply the following MF decoupling to the interactions on the $z$ bond $\rho=\means{jj'}_z$ with $j$ ($j'$) being the A(B)-sublattice site as
\begin{align}
 i a_j b_{j'}i \bar{a}_j \bar{b}_{j'}\simeq -i A b_{j'} \bar{b}_{j'} -iB a_j \bar{a}_j +AB
+i\bar{\Phi} a_j b_{j'} +i\Phi \bar{a}_j \bar{b}_{j'} -\Phi\bar{\Phi}
-i\Theta \bar{a}_j b_{j'} -i\bar{\Theta} a_j \bar{b}_{j'} +\Theta\bar{\Theta}.
\end{align}
Here, we introduce the MFs
\begin{align}
 A=i\means{a_j \bar{a}_j}=\means{2S_j^z},\quad B=i\means{b_{j'} \bar{b}_{j'}}=-\means{2S_{j'}^z},\quad
 \Phi=i\means{a_j b_{j'}},\quad \bar{\Phi}=i\means{\bar{a}_j \bar{b}_{j'}}=\means{\eta_\rho},\quad \Theta=i\means{a_j \bar{b}_{j'}},\quad\bar{\Theta}=i\means{\bar{a}_j b_{j'}},
\end{align}
which are all real and assumed to be site-independent in the present approximation.
 Note that $A$ and $-B$ represent the magnetization in the $z$ direction, and $\bar{\Phi}$ takes $\pm 1$ at $h=0$ corresponding to the local conserved quantity $\eta_\rho$.
 As the sign of $\bar{\Phi}$ does not affect physical quantities, $\bar{\Phi}$ is chosen to be positive in the present calculations.
 Using the Fourier transformation for the Majorana fermions, we obtain the MF Hamiltonian
\begin{align}
 {\cal H}^{\rm MF}=\sideset{}{'}\sum_{\bm{k}} \bm{c}_{\bm{k}}^\dagger {\cal H}_{\bm{k}}^{\rm MF} \bm{c}_{\bm{k}},
\end{align}
where $\bm{c}_{\bm{k}}=\left(a_{\bm{k}},b_{\bm{k}},\bar{a}_{\bm{k}},\bar{b}_{\bm{k}}\right)^{\rm T}$ and the sum $\sideset{}{'}\sum_{\bm{k}}$ is taken for the half of the Brillouin zone so as to avoid the redundancy between $\bm{k}$ and $-\bm{k}$.
${\cal H}_{\bm{k}}^{\rm MF}$ is given by
\begin{align}
 {\cal H}_{\bm{k}}^{\rm MF}=
\begin{pmatrix}
0 &
-iJ(e^{-i k_a}+e^{-i k_b}+\bar{\Phi})e^{i\delta_{\bm{k}}}/2 &
i(JB/2-h) &
iJ\bar{\Theta}e^{i\delta_{\bm{k}}}/2\\
iJ(e^{i k_a}+e^{i k_b}+\bar{\Phi})e^{-i\delta_{\bm{k}}}/2 &
0 &
-iJ\Theta e^{-i\delta_{\bm{k}}}/2 &
i(JA/2+h)\\
-i(JB/2-h) &
iJ\Theta e^{i\delta_{\bm{k}}}/2 &
0 &
-iJ\Phi e^{i\delta_{\bm{k}}}/2\\
-iJ\bar{\Theta}e^{-i\delta_{\bm{k}}}/2 &
-i(JA/2+h) &
iJ\Phi e^{-i\delta_{\bm{k}}}/2 &
0 &
\end{pmatrix}.
\end{align}
Here, $k_a=(k_x+\sqrt{3}k_y)/2$, $k_b=(k_x-\sqrt{3}k_y)/2$, and $\delta_{\bm{k}}=k_y/\sqrt{3}$, where we take the lengths of the primitive translation vectors unity.
The MFs are determined self-consistently through the diagonalization of ${\cal H}_{\bm{k}}^{\rm MF}$.
We note that the MFs $\Theta$ and $\bar{\Theta}$ are always zero in the MF solutions.

\section{Majorana dispersion}

\begin{figure}[t]
 \begin{center}
  \includegraphics[width=0.6\columnwidth,clip]{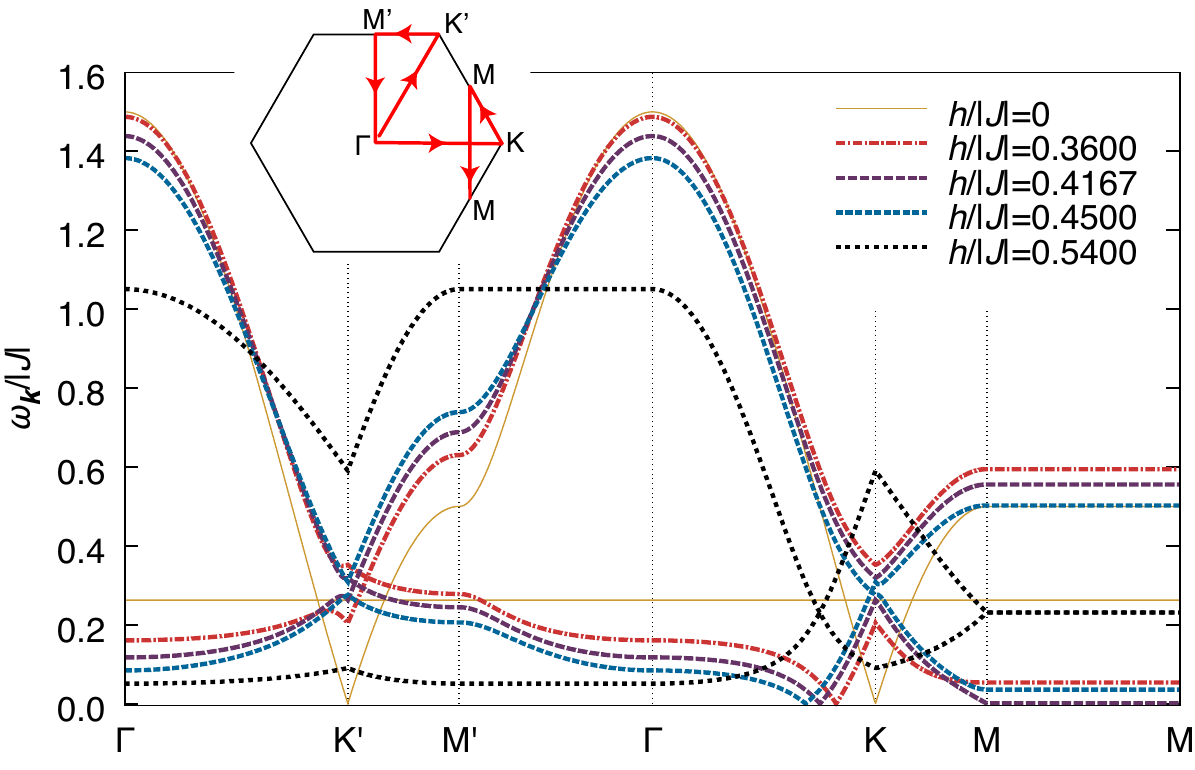}
  \caption{
Dispersion relations corresponding to Fig.~3 of the main text as a function of $\bm{k}$ along the symmetric line shown in the inset.
The data at $h=0$ is also shown.
}
  \label{supp_disp}
 \end{center}
\end{figure}

\begin{figure}[t]
 \begin{center}
  \includegraphics[width=0.6\columnwidth,clip]{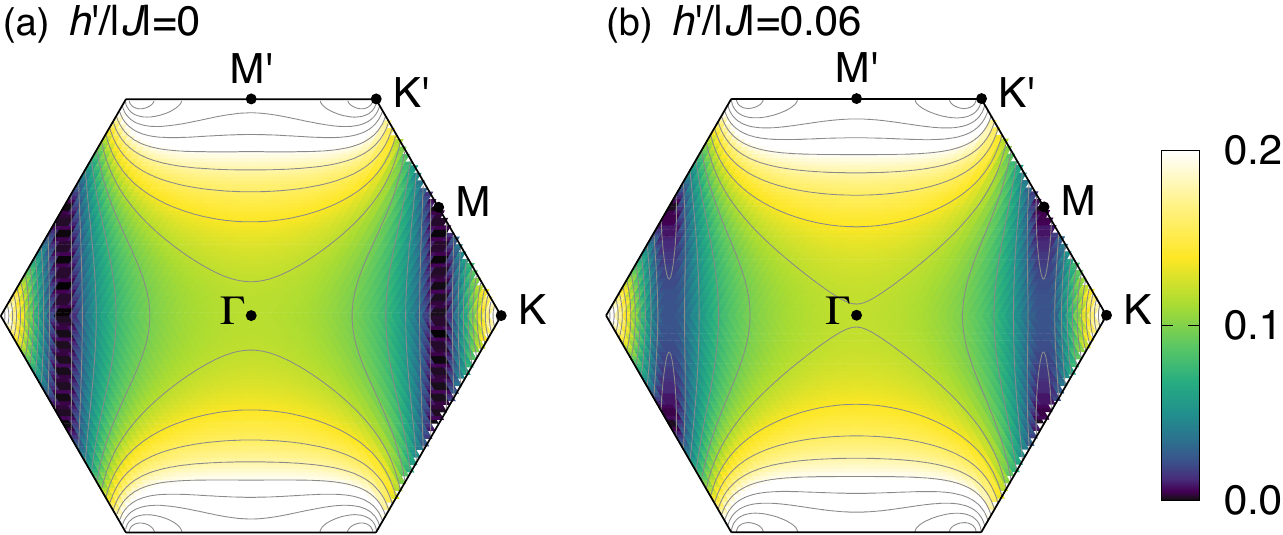}
  \caption{
Dispersions at (a) $h'=0$ and (b) $h'/|J|=0.06$ at the critical field $h=h_{c1}$.
Here, $h'$ is the effective magnetic field defined in the main text.
}
  \label{supp_critical}
 \end{center}
\end{figure}

 We show the Majorana dispersion $\omega_{\bm{k}}$ obtained by the Majorana MF theory in Fig.~\ref{supp_disp}.
 The low-energy part is presented in Fig.~3 of the main text.

 At $h=h_{c1}\simeq 0.4167J$, $\omega_{\bm{k}}$ is zero along the M--M line, which corresponds to the vertical line node shown in Fig.~\ref{supp_critical}(a).
 This structure is lifted by a small effective magnetic field $h'$ introduced in the main text.
 For a nonzero $h'$, the nodal line at $h'=0$ shrinks to a nodal point at the M point as shown in Fig.~\ref{supp_critical}(b).

\section{Low-energy eigenstate of Majorana cones}

 Let us discuss the low-energy physics around the nodal point of the Majorana cone appearing in the Majorana MF solution at $\bm{k}_0=(k_0,0)$.
 Here, $k_0$ is given by
\begin{align}
 \cos \frac{k_0}{2}=-\frac{1}{2\Phi}\left[\bar{\Phi}\Phi+\left(2h/J+A\right)^2\right]\equiv -\frac{\xi}{2},
\end{align}
where we introduce the parameter $\xi$ and assume $B=-A$, meaning that the same magnetic moment appears at the A- and B-sublattice sites.
We have confirmed that this condition is always satisfied in the MF solutions.

By expanding the eigenenergy of ${\cal H}_{\bm{k}}^{\rm MF}$ around $\bm{k}_0$, we obtain the low-energy dispersion,
\begin{align}
 E_{\tilde{\bm{k}}}\sim\sqrt{|v_x|^2 \tilde{k}_x^2+|v_y|^2 \tilde{k}_y^2},
\end{align}
where $\tilde{\bm{k}}=\bm{k}-\bm{k}_0$, and the velocities,
\begin{align}
 v_x={\rm sgn}(k_0)\frac{J\Phi\sqrt{4-\xi^2}}{4(\bar{\Phi}-\Phi-\xi)}, \quad
v_y=\frac{\sqrt{3} J\Phi \xi}{4(\bar{\Phi}-\Phi-\xi)}.\label{eq:2}
\end{align}
 Note that the latter in Eq.~(\ref{eq:2}) indicates that the sign change of $\xi$ results in that of $v_y$, as shown in Figs.~4(a) and~4(b) of the main text.

 The eigenstate corresponding to the eigenenergy $E_{\tilde{\bm{k}}}$ around $\bm{k}_0$ is given by
\begin{align}
\kets{\varphi_{\tilde{\bm{k}}}}
\sim\frac{1}{\sqrt{1+|r_{\tilde{\bm{k}}}|^2}}
\left(1,e^{i\phi_{\tilde{\bm{k}}}-i\delta_{\bm{k}}},
r_{\tilde{\bm{k}}}e^{i\phi_{\tilde{\bm{k}}}},
r_{\tilde{\bm{k}}}e^{-i\delta_{\bm{k}}}
\right)^{\rm T},
\end{align}
as shown in the main text.
 Here, the coefficient $r_{\tilde{\bm{k}}}$ and the phase $\phi_{\tilde{\bm{k}}}$ are explicitly obtained as
\begin{align}
r_{\tilde{\bm{k}}}=
\frac{2h/J+A}{\Phi}
\left(
1-{\rm sgn}[k_0]\frac{2(v_x \tilde{k}_x-i v_y \tilde{k}_y)}{J\Phi}
\right),
\quad
 e^{-i\phi_{\tilde{\bm{k}}}}=\frac{v_y \tilde{k}_y +i v_x \tilde{k}_x} {E_{\tilde{\bm{k}}}}.
\end{align}

\section{Spin-wave theory}

In this section, we briefly introduce the calculation method and the results of the spin-wave (SW) theory.
Here, we start from the state where all spins are fully polarized to the [001] direction.
We apply the Holstein-Primakoff transformation, which is given by
\begin{align}
 S_j^+=\sqrt{2S}\sqrt{1-\frac{\alpha_j^\dagger \alpha_j}{2S}} \alpha_j,\quad S_j^z=S-\alpha_j^\dagger \alpha_j
\end{align}
for the A-sublattice sites and
\begin{align}
 S_j^+=\sqrt{2S}\sqrt{1-\frac{\beta_j^\dagger \beta_j}{2S}} \beta_j,\quad S_j^z=S-\beta_j^\dagger \beta_j
\end{align}
for the B-sublattice sites.
Here, $\alpha_j^\dagger$ and $\beta_j^\dagger$ are the creation operators of bosons.
Using their Fourier transformations $\alpha_{\bm{k}}^\dagger$ and $\beta_{\bm{k}}^\dagger$, the Hamiltonian given by Eq.~(1) of the main text is written as
\begin{align}
 {\cal H}\simeq {\cal H}^{\rm SW}=\sideset{}{'}\sum_{\bm{k}} \bm{\gamma}_{\bm{k}}^\dagger {\cal H}_{\bm{k}}^{\rm SW} \bm{\gamma}_{\bm{k}}-\frac{N}{2}\left\{J S(S+1)+h(2S+1)\right\},
\end{align}
within the linear SW approximation.
Here, we introduce $\bm{\gamma}_{\bm{k}}=\left(\alpha_{\bm{k}},\beta_{\bm{k}},\alpha_{-\bm{k}}^\dagger,\beta_{-\bm{k}}^\dagger\right)^{\rm T}$ and ${\cal H}_{\bm{k}}^{\rm SW}$ is given by
\begin{align}
 {\cal H}_{\bm{k}}^{\rm SW}=
\begin{pmatrix}
JS+h &
-JS(e^{-i k_a}+e^{-i k_b})e^{i\delta_{\bm{k}}}/2 &
0 &
-JS(e^{-i k_a}-e^{-i k_b})e^{i\delta_{\bm{k}}}/2\\
-JS(e^{i k_a}+e^{i k_b})e^{-i\delta_{\bm{k}}}/2 &
JS+h &
-JS(e^{i k_a}-e^{i k_b})e^{-i\delta_{\bm{k}}}/2 &
0\\
0 &
-JS(e^{-i k_a}-e^{-i k_b})e^{i\delta_{\bm{k}}}/2 &
JS+h &
-JS(e^{-i k_a}+e^{-i k_b})e^{i\delta_{\bm{k}}}/2\\
-JS(e^{i k_a}-e^{i k_b})e^{-i\delta_{\bm{k}}}/2 &
0 &
-JS(e^{i k_a}+e^{i k_b})e^{-i\delta_{\bm{k}}}/2 &
JS+h
\end{pmatrix}.
\end{align}
By diagonalizing this via the Bogoliubov transformation, we calculate the magnetization $\means{S^z}$ for the ground state defined by the vacuum of the Bogoliubov bosons.

 As shown in Fig.~1(a) of the main text, in the FM case, $M^z$ shows a similar trend to the results obtained by the other two methods in the whole range of $h$, except for the low-field QSL region.
 The deviation becomes more conspicuous in the AFM case shown in Fig.~1(b) of the main text.
 We note that a spin canted state is realized within the classical MF solution below $h/|J|=1$~\cite{Janssen2017}.
 However, in this state, there exists a flat magnon dispersion at the zero energy due to the frustration inherent to the Kitaev model with bond-dependent interactions.
 This indicates that the canted state is unstable within the linear SW approximation.

\section{Spin correlations in the [001] magnetic field}

\begin{figure}[t]
 \begin{center}
  \includegraphics[width=0.9\columnwidth,clip]{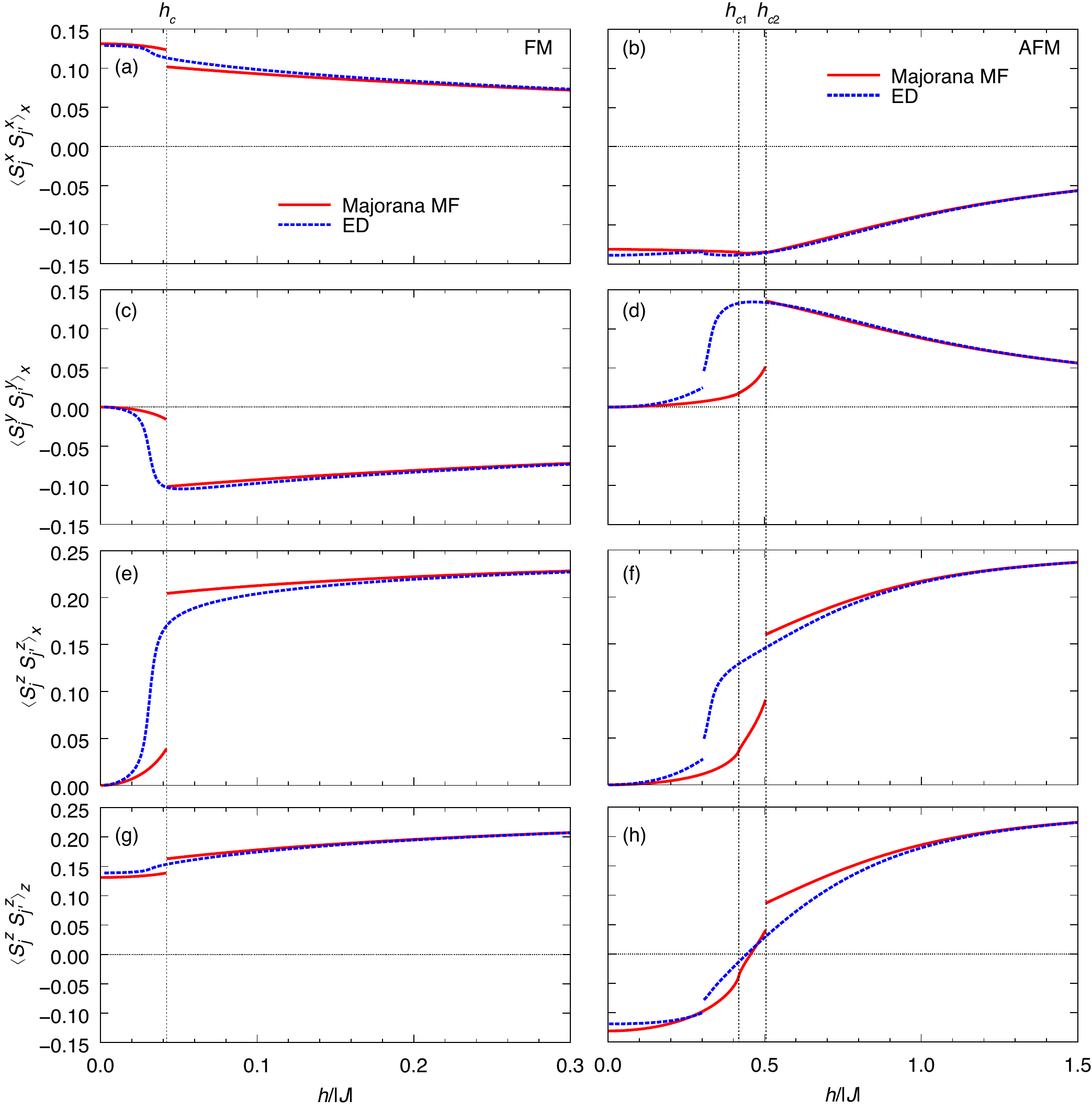}
  \caption{
 Magnetic-field dependence of the spin correlations $\means{S_j^\gamma S_{j'}^\gamma}_{\gamma'}$ on the nearest-neighbor $\gamma'$ bonds:
(a), (c), (e), (g) for the FM Kitaev model and (b), (d), (f), (h) for the AFM Kitaev model.
The results are calculated by the Majorana MF theory and the ED on the 24-site cluster in the inset of Fig.~1(b) in the main text.
}
  \label{supp_ss}
 \end{center}
\end{figure}

 In this section, we compare the results for the spin correlations obtained by the Majorana MF theory with those by the ED in the 24-site cluster.
 Figure~\ref{supp_ss} shows the nearest-neighbor spin correlations as functions of the [001] field $h$ in the FM and AFM Kitaev models.
 Here, $\means{S_j^\gamma S_{j'}^\gamma}_{\gamma'}$ represents the spin correlation of the $\gamma$ component on the $\gamma'$ bond $\means{jj'}_{\gamma'}$.
 In the absence of the magnetic field, the spin correlations have the same amplitude with the opposite signs between the FM and AFM cases; we note that the amplitudes in the ED results are slightly different between the two cases because of the threefold degeneracy in the 24-site cluster.
In the presence of the magnetic field in the [001] direction, the relations $K\equiv\means{S_j^x S_{j'}^x}_{x}=\means{S_j^y S_{j'}^y}_{y}$ and $\bar{K}\equiv\means{S_j^y S_{j'}^y}_{x}=\means{S_j^x S_{j'}^x}_{y}$ are satisfied, and in terms of the Majorana representation, $|K|$ and $|\bar{K}|$ give a measure of the kinetic energy of the Majorana fermions $(a_j,b_{j'})$ and $(\bar{a}_j,\bar{b}_{j'})$, respectively.
 On the other hand, in the present MF approach, it is difficult to evaluate $\means{S_j^x S_{j'}^x}_{z}$ and $\means{S_j^y S_{j'}^y}_{z}$ because these terms involve the string factor appearing in the Jordan-Wigner transformation.
 As shown in Fig.~\ref{supp_ss}, the spin correlations obtained by the Majorana MF theory agree well with those by the ED.
 Particularly, the magnetic field at which the sign change of $\means{S_j^z S_{j'}^z}_{z}$ occurs in the Majorana MF theory is close to that of the ED, as shown in Fig.~\ref{supp_ss}(h).
 Moreover, we find that $|\bar{K}|$ shows a nonmonotonic $h$ dependence in both FM and AFM cases, as shown in Figs.~\ref{supp_ss}(c) and \ref{supp_ss}(d); this quantity grows from zero as increasing $h$ from $h=0$, and takes a peak in the magnitude around the critical fields.

\begin{figure}[t]
 \begin{center}
  \includegraphics[width=0.9\columnwidth,clip]{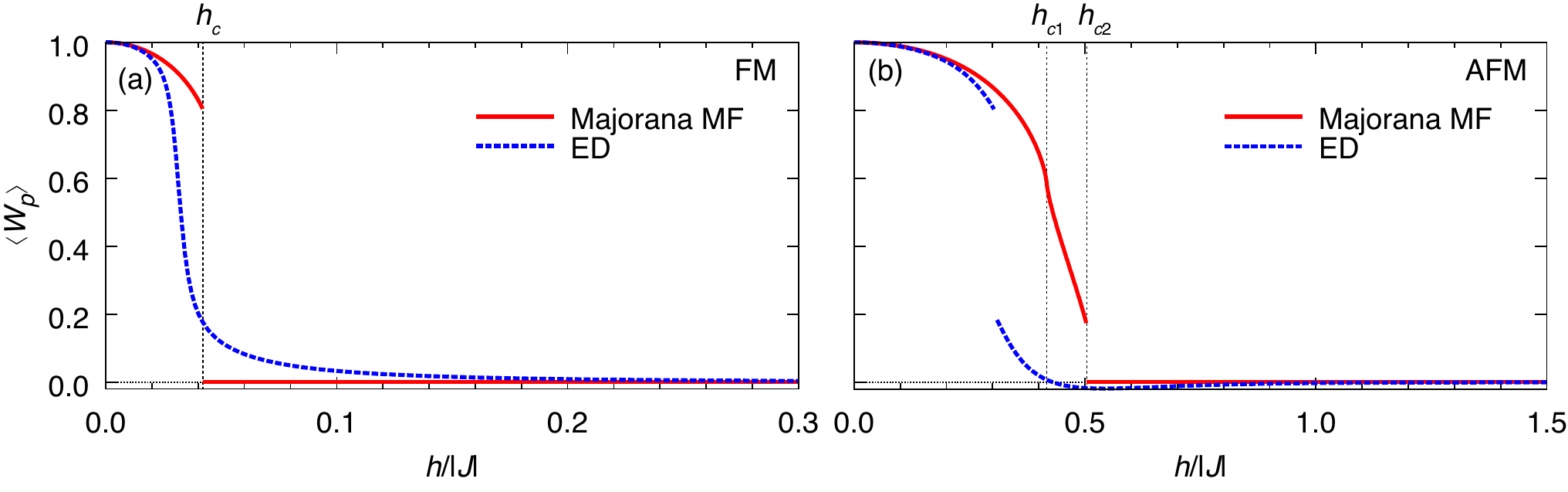}
  \caption{
Magnetic field dependence of $\means{W_p}$ for (a) the FM Kitaev model and (b) the AFM one, which are calculated by the Majorana MF theory and the ED on the 24-site cluster.
}
  \label{supp_w}
 \end{center}
\end{figure}

 We also calculate the quantity $W_p=2^6\prod_{j\in p} S_j^{\bar{\gamma}_j}$ in the hexagon plaquette $p$, where $\bar{\gamma}_j$ stands for the bond index not belonging to the hexagon $p$ among three bonds connected with site $j$.
 This is a local $Z_2$ conserved quantity and takes $+1$ in the absence of the magnetic field~\cite{Kitaev2006}.
 Figures~\ref{supp_w}(a) and~\ref{supp_w}(b) show $\means{W_p}$ in the FM and AFM Kitaev models, respectively.
 For both the two cases, $\means{W_p}$ abruptly decreases around the critical fields while increasing $h$.

\section{Low-energy Majorana eigenstates in the anisotropic Kitaev model}

In this section, we show the low-energy properties of the Kitaev model with bond anisotropy in the absence of the magnetic field, as a reference for the magnetic field effect on the isotropic Kitaev model discussed in the main text.
The model is given by the Hamiltonian in Eq.~(1) with $\bm{h}=0$ in the main text.
In the present calculation, we change $J_z$ while setting $J_x=J_y$.
The ground state was exactly obtained as a free Majorana system with the so-called flux-free configuration, i.e., all $W_p$ defined in the previous section being $+1$~\cite{Kitaev2006}.
This is written by using the Majorana representation used in the main text as
\begin{align}
 {\cal H}=-iJ_x\sum_{\gamma=x,y}\sum_{\mean{jj'}_\gamma}a_j b_{j'}-iJ_z\sum_{\mean{jj'}_z}a_j b_{j'},
\end{align}
where the local conserved quantities $\eta_\rho =i \bar{a}_j \bar{b}_{j'}$ are set to $+1$ on all the $z$ bonds corresponding to the flux-free state.
 By the Fourier transformation, the Hamiltonian is written as
\begin{align}
 {\cal H}=\sideset{}{'}\sum_{\bm{k}}
\begin{pmatrix}
a_{\bm{k}}^\dagger & b_{\bm{k}}^\dagger
\end{pmatrix}
{\cal H}_{\bm{k}}
\begin{pmatrix}
a_{\bm{k}}\\ b_{\bm{k}}
\end{pmatrix},
\end{align}
with
\begin{align}
 {\cal H}_{\bm{k}}=
\begin{pmatrix}
0 &
-iJ_x(e^{-i k_a}+e^{-i k_b}+ J_z/J_x)e^{i\delta_{\bm{k}}}/2 \\
iJ_x(e^{i k_a}+e^{i k_b}+J_z/J_x)e^{-i\delta_{\bm{k}}}/2 &
0
\end{pmatrix}.
\end{align}
 The eigenvalue of ${\cal H}_{\bm{k}}$ in the positive energy sector is obtained by the diagonalization as
\begin{align}
 E_{\bm{k}}=J_x\sqrt{\cos^2 (k_x/2)+(J_z/J_x)^2/4+(J_z/J_x)\cos (k_x/2) \cos(\!\sqrt{3}k_y/2)}.
\end{align}
 From this representation, we find that the Majorana cones appear with the nodal points located at $\bm{k}=\bm{k}_0=(k_0,0)$, where
\begin{align}
 \cos\frac{k_0}{2}=-\frac{J_z}{2J_x}.
\end{align}
 Corresponding to $\xi$ discussed in the main text, we introduce the parameter $\xi$ so that $\xi=-2\cos(k_0/2)$.
 In this case, therefore, $\xi$ is nothing but the anisotropy of the Kitaev interactions, namely,
\begin{align}
\xi = \frac{J_z}{J_x}.
\end{align}
Expanding $E_{\bm{k}}$ around $\bm{k}_0$, we obtain the low-energy behavior as
$E_{\tilde{\bm{k}}}\sim \sqrt{|v_x|^2 \tilde{k}_x^2+|v_y|^2 \tilde{k}_y^2}$, where $\tilde{\bm{k}}=\bm{k}-\bm{k}_0$, and the velocities  given by
\begin{align}
 v_x={\rm sgn}(k_0)\frac{J_x\sqrt{4-\xi^2}}{4}, \quad
v_y=\frac{\sqrt{3} J_x\xi}{4}.
\end{align}
 These expressions are the same as those in Eq.~(\ref{eq:2}) except for the factor of $\frac{\Phi}{\bar{\Phi}-\Phi-\xi}$, which appears due to the renormalization originating from the mixing between the Majorana fermions $(a_j,b_{j'})$ and $(\bar{a}_j,\bar{b}_{j'})$.
The low-energy wavefunction is calculated as
$\kets{\varphi_{\tilde{\bm{k}}}}
=\left(1,e^{i\phi_{\tilde{\bm{k}}}-i\delta_{\bm{k}}}
\right)^{\rm T}$
for the basis of ($a_{\bm{k}}, b_{\bm{k}}$) with $e^{-i\phi_{\tilde{\bm{k}}}}=(v_y \tilde{k}_y +i v_x \tilde{k}_x)/E_{\tilde{\bm{k}}}$.
 This is also equivalent to Eq.~(2) of the main text in the Majorana fermion sector of $(a_j,b_{j'})$.
 Therefore, the topology of the Majorana fermion band in the anisotropic Kitaev model is altered by changing the sign of $\xi$, namely, the anisotropy $J_z/J_x$, in a similar manner to the AFM isotropic Kitaev model in the [001] field discussed in the main text.

\end{document}